\documentclass{aastex}
\usepackage{emulateapj5}
\usepackage{apjfonts}

\input{epsf.sty}

\newcommand\Msun{{M$_\odot$}}
\newcommand\rmin{{r_{\rm min}}}

\slugcomment{Accepted for publication in ApJ Letters}

\begin{document}

\title{Thermal Comptonization and Disk Thermal Reprocessing in NGC 3516}

\author{James Chiang\altaffilmark{1,2} and Omer Blaes\altaffilmark{3}}
\altaffiltext{1}{NASA/GSFC, Code 661, Greenbelt MD 20771}
\altaffiltext{2}{Joint Center for Astrophysics/Physics Department, 
         University of Maryland, Baltimore County, Baltimore MD 21250}
\altaffiltext{3}{Department of Physics, University of California, 
       Santa Barbara CA 93106-9530}

\setcounter{footnote}{0}

\begin{abstract}
We present an application of the thermal Comptonization/disk
reprocessing model recently proposed by Zdziarski, Lubi\'nski, \&
Smith.  We show that the absence of strong optical variations in the
presence of strong concurrent X-ray variations, similar to those found
by HST/RXTE monitoring observations of NGC 3516, can be explained by
changing the geometry of the Comptonizing plasma rather than the
accretion disk itself.  The total X-ray luminosity of the Comptonizing
plasma must decrease as its spatial extent increases.  In contrast,
the disk inner radius must be roughly fixed in order not to produce
optical/ultraviolet color variations stronger than observed.  By
including emission due to internal viscous dissipation in the disk, we
can roughly match the optical and X-ray flux levels and variability
amplitudes seen from NGC 3516 during the HST/RXTE campaign.
\end{abstract}

\keywords{galaxies: active --- galaxies: individual (NGC 3516)
--- galaxies: Seyfert --- X-rays: galaxies}

\section{Introduction}
A ubiquitous feature of the optical/UV emission from Type 1 Seyfert
galaxies (Sy1s) is the highly correlated variability of the continuum
flux across a wide range of wavelengths (e.g., Peterson et al.\ 1998).
When this behavior was first observed during the 1989--1990 broad line
region reverberation mapping campaign for NGC~5548 (Peterson et al.\
1991; Clavel et al.\ 1991), it was immediately recognized that the
upper limits for the relative lags between the optical and UV
wavebands were too small for the variability to have been mediated by
hydrodynamic processes in the putative accretion disk --- the
differences in the characteristic radii for the emission in the
respective wavebands were too large.  In fact, the upper limits for
the interband lags of $\la 2$\,days required signal speeds of $\ga 0.1
c$ which could not have occurred within the disk itself.  This fact
and the observation that the optical/UV continuum became bluer as the
flux increased led to the conclusion that the continuum variations
were due to thermal reprocessing in the disk of higher energy
radiation produced by a single varying source, such as the X-ray
source posited to exist at the disk's inner regions, near the central
black hole (Krolik et al.\ 1991; Courvoisier \& Clavel 1991;
Collin-Souffrin 1991).  Subsequent observations of other Sy1s showed
that highly correlated broad band optical/UV variability is generic to
these objects (e.g., Courvoisier \& Clavel 1991).

Recently, the thermal reprocessing picture has been challenged by
simultaneous optical, UV, and X-ray observations of Sy1s.  The first
object so observed was NGC~7469 by the IUE and RXTE satellites over
$\sim 30$\,days in 1997 (Nandra et al.\ 1998).  Although the UV and
X-ray light curves showed some similar variability characteristics
both in amplitude and time scales, these light curves were not nearly
as well correlated as would be expected if the UV emission was due to
thermal reprocessing.  Naively, the variations in the UV should follow
those in the X-rays with some characteristic time delay, but the
maxima in the UV light curves preceded similarly shaped maxima in the
X-ray light curve by $\sim 4$\,days, while minima in both light curves
occurred nearly simultaneously.

The 1998 observations of NGC 3516 by HST, ASCA, and RXTE seem to be
equally troubling for the thermal reprocessing model (Edelson et al.\
2000).  These observations consisted of intensive monitoring in
optical, UV, and X-ray bands over 3~days and so were able to probe
correlations on far shorter time scales than any previous set of
observations.  The X-rays showed very strong variations ($\sim 60$\%
peak-to-peak), while the changes in the optical continuum were much
smaller ($\sim 3$\%).  Because of the accurate relative photometry
afforded by the HST/STIS CCDs, the measurement uncertainties were
sufficiently small to reveal significant variability that was strongly
correlated across the optical bands.  Hence, the original motivation
for the thermal reprocessing model was still present, but as for NGC
7469, the X-ray and optical flux variations were not strongly
correlated on any relevant time scale.

More recently, Nandra et al.\ (2000) have re-examined the 1997
observations of NGC~7469 and found that while the relatively narrow
band 2--10\,keV X-ray flux wasn't well-correlated with the UV flux the
spectral index of the X-ray emission was.  This led them to suggest
that the UV emission is actually correlated with the {\em bolometric}
X-ray flux.  In this Letter, we expand upon that idea and describe
what is required for the multiwaveband phenomenology exhibited by NGC
3516 to be due to disk thermal reprocessing in the context of a
thermal Comptonization model for the X-ray emission.

\section{X-ray Thermal Reprocessing in the Sphere+Disk Geometry}

Most of the important details of the model are described in Zdziarski,
Lubi\'nski, \& Smith (1999; hereafter ZLS) which the interested reader
is encouraged to consult.  We will proceed by summarizing the
significant features and results from that work.  This model was
proposed by ZLS to explain the apparent correlation between the X-ray
spectral index, $\Gamma$, and the relative magnitude, $R$, of the
so-called Compton reflection hump seen in the hard X-ray spectra of
Sy1s and X-ray binaries.  ZLS argued that this correlation results
from the reprocessing and Comptonizing interactions between the disk
and the hot plasma believed to produce the X-ray continuum.  Expanding
upon the work of previous authors (e.g., Poutanen, Krolik, \& Ryde
1997; Zdziarski et al.\ 1998), ZLS noted that the $R$-$\Gamma$
correlation could arise in a geometry in which the Comptonizing plasma
occupies a spherical region of radius $r_s$ centered on the black
hole.  The accretion disk has an inner radius $\rmin$ which penetrates
this sphere at its equator by varying amounts.  This varying
penetration could be due to changes in $r_s$, $\rmin$, or both.
Smaller values of $\rmin/r_s$ result in a larger fraction of the X-ray
photons from the sphere being intercepted by the disk.  The covering
factor, $R$, for Compton reflection and thermal reprocessing is
larger, and a larger fraction of the disk soft photons re-enter the
plasma precipitating additional cooling via Compton losses and thereby
reducing the temperature of the plasma.  This results in a softer
X-ray spectrum and a larger value for the photon index $\Gamma$.
Enforcing energy balance under the assumption that the disk emits only
reprocessed radiation, ZLS derived a $R$-$\Gamma$ relation which looks
very similar to the observed correlations.

We follow ZLS by assuming that the region of Comptonizing plasma is
homogeneous and optically thin and that we can neglect internal
viscous dissipation in the disk, at least initially.  Reprocessing of
incident X-rays by a razor-thin disk then gives a radial effective
temperature distribution
\begin{equation}
T(r) = \left[\frac{3}{16 \pi^2
             \sigma_B}\frac{L_x}{r_s^2}\,h\left(\frac{r}{r_s}\right)
             (1 - a)\right]^{1/4}.
\label{T(r)}
\end{equation}
Here $L_x$ is the total luminosity of the Comptonizing plasma,
$\sigma_B$ is the Stefan-Boltzmann constant, and $a$ is the disk
albedo.  The dimensionless function $h(r/r_s)$ is given by Eq. 1 of
ZLS.  

Assuming local blackbody emission and that the optical portion of the
observed spectrum originates at large radii, we may use the asymptotic
form $h(r/r_s \gg 1) \rightarrow (\pi/4)(r_s/r)^3$ in Eq.~\ref{T(r)}
to obtain the disk spectrum
\begin{equation}
F_\nu \propto \nu^3\int_\rmin^\infty r\,dr
     \frac{1}{\exp\left(E_\nu r^{3/4}/k_B {\cal A}\right) - 1},
\label{Fopt}
\end{equation}
where $T = {\cal A} r^{-3/4}$, ${\cal A} \equiv [3 L_x r_s
(1-a)/64\pi\sigma_B]^{1/4}$, and $E_\nu$ is the photon energy.
Neglecting variations in the disk albedo and assuming that $\rmin$ is
fixed on the time scales of interest, a constant optical flux implies
a constant value of ${\cal A}$.  In the presence of variations of the
X-ray luminosity, this requires $L_x \propto r_s^{-1}$.

For $r \simeq r_s$, i.e., at higher frequencies, it turns out that the
large radius limit of $h(r/r_s)$ is a surprisingly good approximation
for evaluating the disk spectrum since it enters as the 1/4-power in
$T$, and at $r/r_s = 1$, the function takes the value $h(1) = 4/3$
(ZLS), and we have $(\pi/4)^{1/4} \approx (4/3)^{1/4} \approx 1$.
Hence, for $r_s \simeq \rmin$, adopting the asymptotic form for
$h(r/r_s)$ throughout and setting $L_x \propto r_s^{-1}$, all
$r_s$-dependence drops out of Eq.~\ref{Fopt}, and in this
approximation, the resulting disk spectrum has a constant shape and
magnitude for different values of $r_s$.  The exact expression for
$h(r/r_s)$ flattens for $r < r_s$, so when $\rmin < r_s$, the disk
spectrum predicted by the full calculation will be somewhat softer,
but not appreciably at the wavelengths which have actually been
observed.

Given the X-ray luminosity $L_x$, we estimate the X-ray spectrum
produced by thermal Comptonization in the sphere.  For this we use a
second result from ZLS which relates the amplification factor for
Comptonization, $A$, to the ratio $\rmin/r_s$ (ZLS, Eq.~2).  Following
ZLS, we use the empirical result of Beloborodov (1999a) which gives
the X-ray spectral index as a function of $A$: $\Gamma \approx 2.33 (1
- A)^{-1/10}$.  To determine the overall magnitude of the X-ray
spectrum, we use another empirical result of Beloborodov (1999a) to
estimate the temperature of the Comptonizing plasma, $\Gamma \approx
(9/4)y^{-2/9}$, where the Compton parameter is given by $y = 4
(\theta_e + 4\theta_e^2) \tau(\tau + 1)$ and $\theta_e = k_B T_e/m_e
c^2$.  We assume a typical value for the Thomson depth, $\tau = 1$,
and solve the preceding equations to find the electron temperature
$T_e$.  We model the X-ray continuum as a power-law with an
exponential roll-over, $L_E \propto (E/E_c)^{1 - \Gamma}
\exp(-E/E_c)$, where the cut-off energy is taken to be $E_c = 2 k_B
T_e$ appropriate for a Thomson depth $\tau \sim 1$.  The
normalization, $L_0$, of the X-ray spectrum is given by
\begin{equation}
L_x = L_0 \int_{E_{\rm min}/E_c}^\infty 
           (E/E_c)^{1-\Gamma} \exp(-E/E_c) d(E/E_c).
\end{equation}
For $\Gamma < 2$, the precise value of the minimum photon energy is
not very important so we have set it at a nominal value of $E_{\rm
min} = 0.01$\,keV.

\section{Comparisons to Observations}

We have computed thermal Comptonization and disk reprocessing spectra
in this model for NGC 3516.  The X-ray luminosity $L_x$ is found by
fitting the disk spectrum to the mean optical fluxes measured by HST
(Edelson et al.\ 2000).  We fix the disk inner radius at $\rmin =
3\times10^{13}$\,cm and plot the resulting spectral energy
distributions (SEDs) in Figure~\ref{SEDs} for values of the sphere
radius in the range $r_s = 1.5$--$5 \times 10^{13}$\,cm (solid
curves).  As claimed, the calculated disk spectrum is strikingly
constant, with only relatively small variations in the bluer parts of
the spectrum.  For comparison, we plot the mean fluxes measured by
HST.  Given the crudeness of this model, the quality of the fit is
satisfactory even for the UV flux.\footnote{This datum was not
included in the fit for $L_x$ since the UV flux was only monitored
over a significantly shorter period than the optical data due to SAA
passages.  In addition, the UV data point, at $\sim 1350$\AA, is in a
part of the spectrum which may be contaminated by the broad wings of
the Si{\sc iv} emission line (see Edelson et al.\ 2000, Fig.~3).
Unfortunately, similar contamination problems may exist for the
optical data as well (see Edelson et al.\ 2000, Fig.~4).}

Since we have thus far neglected local viscous dissipation in the
disk, the black hole mass does not enter into the calculation of the
disk emission.  However, the value of the disk inner radius can, in
principle, be constrained by the optical and UV data.  The dashed
curve in Figure~\ref{SEDs} is the SED obtained for $\rmin = r_s = 6
\times 10^{13}$\,cm.  This shows that while the disk spectrum is
relatively insensitive to factor $\sim 3$ changes in $r_s$ for a given
optical flux, comparably sized changes in $\rmin$ produce strong color
variations in the UV.  Therefore, in order to account for the UV flux,
the disk inner radius cannot be too much larger than $\rmin = 3 \times
10^{13}$\,cm.  For a central black hole mass of $3 \times 10^7$\,\Msun
(e.g., Edelson \& Nandra 1999), this value for the disk inner radius
is roughly consistent with innermost radius implied by ASCA
observations of the broad Fe\,K$\alpha$ emission line (Nandra et al.\
1999).

Despite the fact that the optical/UV continuum barely changes as a
function of $r_s$, the X-ray spectrum varies drastically.  In the
2--10\,keV band in particular, the flux changes by factors of $\sim 2$
for the spectra shown.  To relate this result more concretely to the
X-ray observations of NGC~3516, we plot in Figure~\ref{Gamma_vs_flux}
the X-ray spectral index versus 2--10\,keV flux for a family of curves
corresponding to different optical fluxes with the position on any
given curve being parameterized by the value of $r_s$.  We propose
that the HST/ASCA/RXTE monitoring observations of NGC 3516 can be
explained in this model by changes in $r_s$ and accompanying changes
in $L_x$ so that $L_x \propto r_s^{-1}$ is approximately maintained,
and the source takes a path in the $\Gamma$-$F_{\rm 2-10}$ plane which
lies nearly along a curve of constant optical flux.  Any significant
variations in the optical/UV continuum will correspond to deviations
from the constant flux curve and
therefore from the $L_x \propto r_s^{-1}$ relation.  Furthermore,
the changes associated with such departures will be broad band and
will show the same color variations that have been previously
associated with thermal reprocessing.

For comparison, we have plotted in Figure~\ref{Gamma_vs_flux} values
of $\Gamma$ and $F_{\rm 2-10}$ which we have obtained from fitting the
publicly archived RXTE NGC 3516 data for the observations in question.
Although this simple model overpredicts the X-ray flux by a factor
$\sim 4$ for the observed range of spectral indices, the correlation
of spectral index and flux given by the model is strikingly similar to
that of the data.  The flux discrepancy could be reduced somewhat by
selecting a smaller value for the Thomson depth $\tau$ since for a
given spectral index (and $y$-parameter) that will imply a larger
value of $T_e$ and more of the X-ray luminosity can be ``hidden'' at
higher energies.  Alternatively, a large reduction can be obtained by
including contributions to the disk emission from local viscous
dissipation.  This corresponds to adding a term $\propto
(GM\dot{M}/r^3)(1 - (r_I/r)^{1/2})$ inside the square brackets in the
expression for the disk temperature (Eq.~\ref{T(r)}) where $r_I = r_g$
for a maximal Kerr metric and $r_I = 6r_g$ for a Schwarzschild metric.
Since the viscous dissipation term is independent of $r_s$, it acts
essentially as a constant offset for the required value of the X-ray
luminosity, and $L_x \propto r_s^{-1}$ is still needed to maintain a
constant optical flux.  In order to match the observed X-ray fluxes,
roughly 1/4 of the total disk luminosity must be due to thermal
reprocessing with the remaining 3/4 being due to viscous dissipation.
Assuming $M = 3 \times 10^7$\,\Msun\ and $r_I = r_g$, we have solved
for the energy balance condition and find that an accretion rate of
$\dot{M} \simeq 3.6 \times 10^{-3}$\,\Msun~yr$^{-1}$ is required to
match the mean X-ray flux.  This solution is plotted as the thick
solid curve in Figure~\ref{Gamma_vs_flux}.

Although the energy balance condition we have used only restricts the
properties of the Comptonizing plasma, {\em global} energy balance is
roughly satisfied by this solution.  The bolometric X-ray luminosity
which gives the observed mean
X-ray flux and spectral index is $L_x \simeq 7.0\times
10^{43}$\,erg~s$^{-1}$, while the above accretion rate implies a disk
luminosity of $L_{\rm disk} \simeq 3.3\times 10^{43}$\,erg~s$^{-1}$
due solely to viscous dissipation.  For a disk extending all the way
to $r_I = r_g$ (rather than being truncated at $r_s$), the total
implied luminosity is $L_{\rm tot}
\simeq 1.0 \times 10^{44}$erg~s$^{-1}$, and we have
\begin{equation}
L_{\rm tot} \approx L_x + L_{\rm disk}.
\label{global_energy_balance}
\end{equation}
This would be expected if the accretion luminosity that would
otherwise emerge as local black body emission from the inner regions
of the untruncated disk were instead partitioned to the heating of the
Comptonizing plasma.  We note that we have arbitrarily set the Thomson
depth to $\tau = 1$ in computing this solution.  By fine tuning the
values of $\tau$ and $\dot{M}$, we find that we can satisfy the global
energy balance condition arbitrarily well while simultaneously
matching the observed mean X-ray flux, spectral index and optical
continuum levels.

\section{Discussion and Conclusions}

The fact that an approximately constant optical/UV continuum can in
principle result from thermal reprocessing of a highly variable X-ray
flux is connected with the fact that the radial distribution of the
flux incident upon the disk is well approximated by a power-law, even
very close to the innermost disk radius.  As we showed above, the
constant flux constraint, in conjunction with the assumed constancy of
the disk inner radius and disk albedo, causes the radius of the
Comptonizing sphere to drop out of the integral describing the disk
spectrum.  However, this constraint does not demand that the shape of
the incident flux distribution be a power-law.  The only requirement
is that the relevant physical parameters affecting the incident flux
can be considered independently from the expression which describes
the {\em shape} of the distribution.  For example, consider the model
proposed by Beloborodov (1999b) in which the Comptonizing medium
consists of a blob of plasma which is at some height $r_h$ above the
disk.  The blob moves with mildly relativistic velocities either
vertically away from or towards the underlying accretion disk, and the
effective reflection fraction is controlled by relativistic aberration
of the flux from the blob.  Using the formulae in Beloborodov (1999b),
the radial distribution of the flux incident upon the disk is given by
\begin{equation}
F_{\rm inc} \propto \frac{L_x}{r_h^2\gamma^4}
            \left[\frac{({\tilde r}^2 + 1)^{1/2}}
                       {\beta + ({\tilde r}^2 + 1)^2}\right],
\end{equation}
where $\beta$ is the blob velocity in units of $c$, $\gamma = (1 -
\beta^2)^{-1/2}$, and ${\tilde r} = r/r_h$ is the disk radius in units
of the blob height.  For the range of blob velocities required by this
model to reproduce the observed $R$-$\Gamma$ correlation, $\beta \sim
0.3$--0.7, the factor in the square brackets which gives the shape of
the flux distribution is largely insensitive to the exact value of
$\beta$.  Therefore, since the lower limit of the radial flux integral
can be taken to be zero (cf.\ Eq.~\ref{Fopt}), a nearly constant
optical flux in this model can be obtained if $L_x \propto
r_h^2\gamma^4$.

Although neither of the aforementioned models offer any reason for the
luminosity scalings which seem to be required by the NGC 3516
observations, any plausible explanation would still have to contend
with why NGC 7469 and probably other Sy1s violate it.  Therefore, we
will not attempt to posit a physically motivated rationale for the
specific relationship between X-ray luminosity and geometry required
for any particular object, but a cursory examination of the important
time scales may provide some insights.  For NGC 3516, the typical
variability time scale exhibited by the 2--10 keV light curve is $\sim
20$--30\,ks (Edelson et al.\ 2000).  Assuming a radial Thomson depth
$\tau = 1$ and $r_s \simeq \rmin$, the cooling time is very short:
$t_{\rm cool} \sim N_e k_B T_e/L_x \la 10$\,s, where $N_e = (4\pi
r_s^3/3)(\tau/r_s \sigma_T)$ is the total number of electrons in the
sphere with mean temperature $T_e \sim 10^9$K.  Unfortunately, the
thermal expansion time for a sphere of this size is fairly long,
approximately $t_{\rm exp} \sim 100$\,ks.  This assumes that the
protons and the electrons are at the same temperature.  If we have a
two-temperature plasma in which the protons possess a virial
temperature of $\sim 10^{12}$\,K, then the expansion time scale is
reduced substantially.  Alternatively, this could also be achieved if
we have a pair plasma instead.

We have assumed for simplicity that the thermal emission from the disk
is locally blackbody.  Detailed modeling of bare disk atmospheres
appropriate for Sy1s indicates that this is a moderately reasonable
approximation in the optical, but significant deviations can occur at
higher frequencies (e.g., Hubeny et al. 2000, 2001).  This appears to
remain true for disks powered exclusively by external X-ray
illumination (Sincell \& Krolik 1997).  Our treatment of the
reprocessed flux in the optical is therefore probably fairly good, but
the full seed photon spectrum illuminating the Comptonizing plasma
will not be very accurate.  More detailed modeling of the disk
spectrum will likely affect the specific values we deduce for the disk
inner radius, X-ray luminosity, and Comptonizing sphere radius, but it
will not change our general conclusions.

The existing simultaneous optical, UV, and X-ray data plus the global
energy balance condition (Eq.~\ref{global_energy_balance}) provide
just enough information to infer unique values for the five model
parameters, $\rmin$, $r_s$, $T_e$, $\tau$, and $\dot{M}$, required to
describe the entire spectral energy distribution.  Therefore, a
crucial test of this general scheme for linking thermal Comptonization
and thermal reprocessing will be provided by simultaneous optical, UV,
and {\em broad band} X-ray monitoring observations which include
reliable measurements of the thermal roll-over at energies $\ga
100$\,keV.  These latter measurements are necessary for making
estimates of the X-ray luminosity $L_x$ which can then be compared
with the luminosities required by the model.  Forthcoming {\em
INTEGRAL} observations of Sy1s will provide some of the necessary
spectral coverage to address this issue, but what is really required
is a hard X-ray/soft gamma-ray telescope which can measure the X-ray
spectrum and thermal roll-over on time scales comparable to the
shortest variability time scales seen from these objects.

\acknowledgements
We thank Andzrej Zdziarski and the anonymous referee for helpful
comments.  JC was partially supported by NASA ATP grant NAG 5-7723
during the course of this work.  OB acknowledges support from NASA
grant NAG 5-7075 and NSF grant PHY 99-07949.  This research has made
use of data obtained from the High Energy Astrophysics Science Archive
Research Center (HEASARC), provided by NASA's Goddard Space Flight
Center.

\clearpage
\centerline{
\parbox[t]{6in}{\hfil\epsfxsize=4in\epsfbox{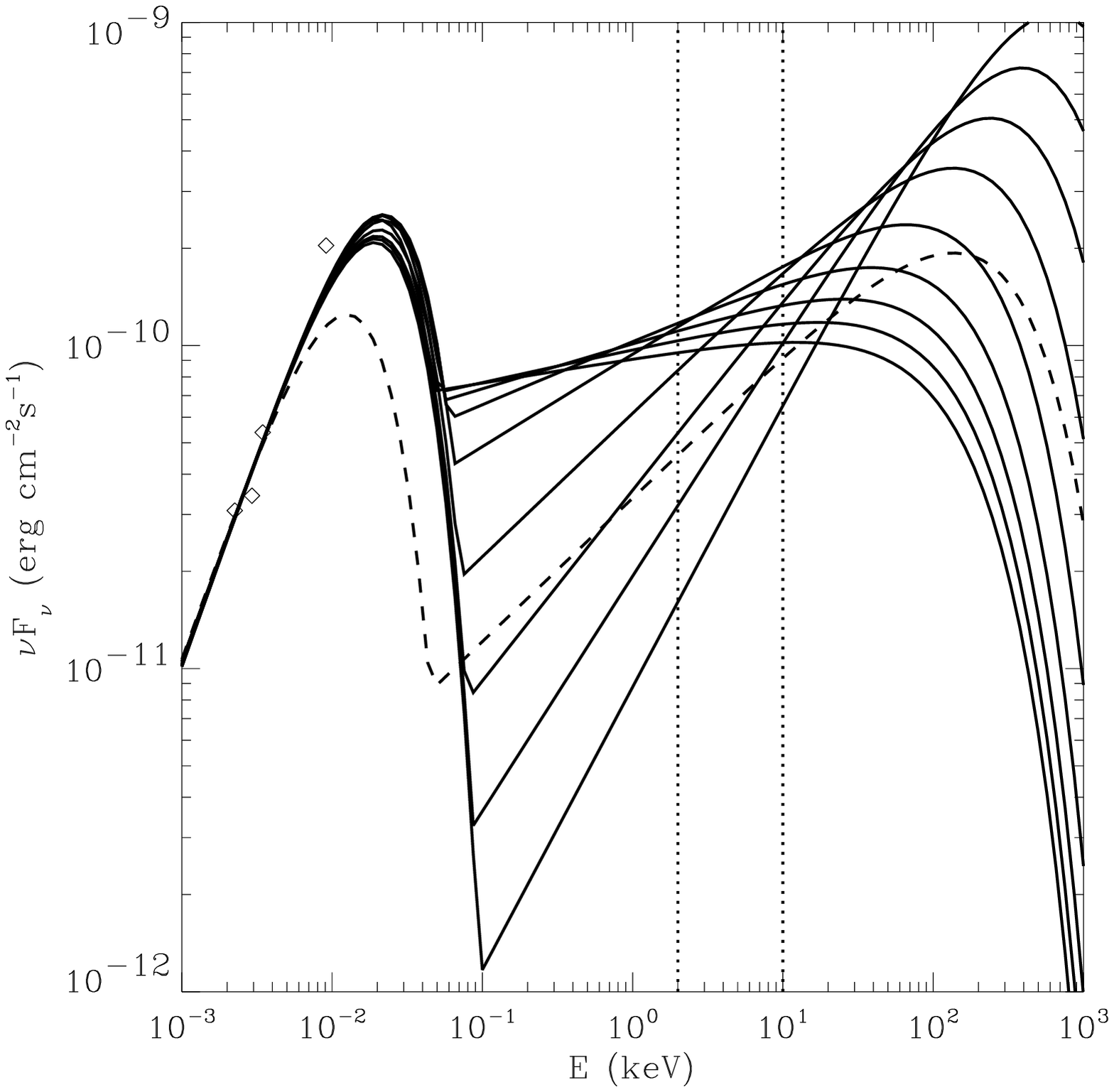}\hfil
\figcaption{Model
SEDs which have been fit to the HST optical fluxes for NGC 3516.  For
the solid curves, the disk inner radius is set at $\rmin = 3\times
10^{13}$\,cm, the Thomson depth of the Comptonizing sphere is $\tau =
1$, and the radius of the sphere takes values $r_s = 1.5$--$5 \times
10^{13}$\,cm.  The smaller values of $r_s$ correspond to the harder
X-ray spectra.  The dashed curve is the SED corresponding to $\rmin =
r_s = 6 \times 10^{13}$\,cm.  The dotted lines indicate the 2--10\,keV
band.
\label{SEDs}\vspace{0.1in}}
}}
\centerline{
\parbox[t]{6in}{\hfil\epsfxsize=4in\epsfbox{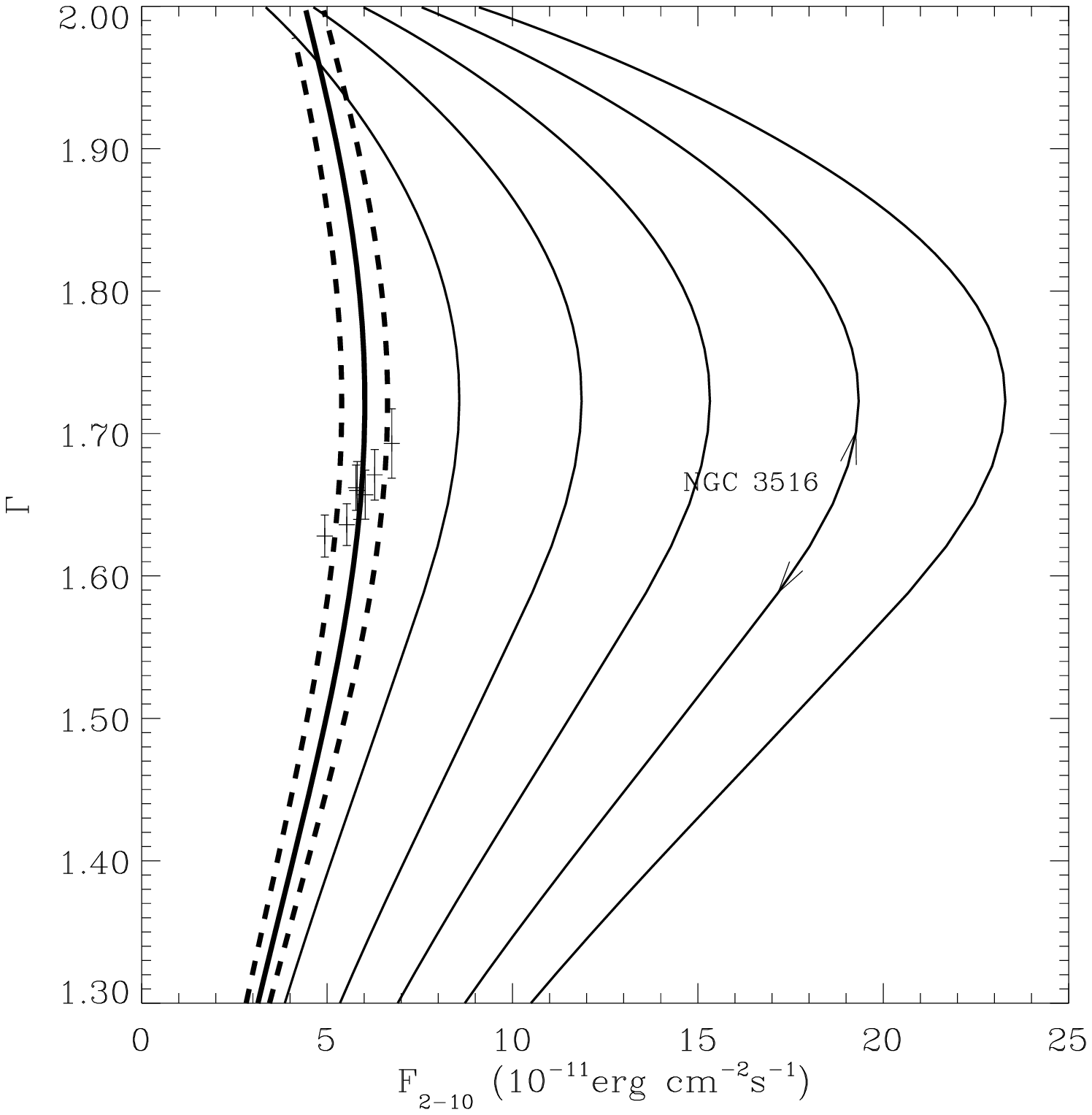}\hfil
\figcaption{X-ray spectral index versus 2--10\,keV flux for families
of curves each of which produces a different constant optical
continuum level.  The rightmost curve represents the model
calculations shown in Figure~\ref{SEDs} and the next four curves to
its left have been computed for 87.5, 75, 62.5, and 50\% of the mean
optical flux level for NGC 3516.  The X-ray emission can vary
significantly and still produce a nearly constant optical/UV continuum
if the source follows a path along one of these curves such as the one
labeled ``NGC 3516''.  The thick solid curve has been fit to the full
optical flux, but emission due to internal viscous dissipation in the
disk has been included to reduce the required X-ray luminosity.  The
neighboring dashed curves bound the region of $\pm 1.5$\% optical
variations.  The data points with error bars are from spectral fits to
the archived 1998 RXTE data for NGC 3516.
\label{Gamma_vs_flux}\vspace{0.1in}}
}}


\begin{thebibliography}{}
\bibitem{B99a} Beloborodov, A. M. 1999a, in High Energy Processes in
   Accreting Black Holes, eds. J. Poutanen and R. Svensson, (San
   Francisco: Astronomical Society of the Pacific), 295
\bibitem{B99b} Beloborodov, A. M. 1999b, ApJ, 510, L123
\bibitem{C91} Clavel, J., et al.\ 1991, ApJ, 366, 64
\bibitem{CS91} Collin-Souffrin, S. 1991, A\&A, 249, 344
\bibitem{CC91} Courvoisier, T. J.-L., \& Clavel, J. 1991, A\&A, 248, 389
\bibitem{E99} Edelson, R. E., \& Nandra, K. 1999, 514, 682
\bibitem{E00} Edelson, R. E., et al. 2000, ApJ, 534, 180
\bibitem{H00} Hubeny, I., Agol, E., Blaes, O., \& Krolik, J. H. 
   2000, ApJ, 533, 710
\bibitem{H01} Hubeny, I., Blaes, O., Krolik, J. H., \& Agol, E. 
   2001, ApJ, in press (astro-ph/0105507)
\bibitem{K91} Krolik, J. H. et al.\ 1991, ApJ, 371, 541
\bibitem{N98} Nandra, K., et al.\ 1998, ApJ, 505, 594
\bibitem{N99} Nandra, K., George, I. M., Mushotzky, R. F., Turner, T. J.,
   \& Yaqoob, T. 1999, ApJ, 523, L17
\bibitem{N00} Nandra, K., et al.\ 2000, ApJ, 544, 734
\bibitem{P91} Peterson, B. M., et al.\ 1991, ApJ, 368, 119
\bibitem{P98} Peterson, B. M., et al.\ 1998, PASP, 110, 660
\bibitem{P97} Poutanen, J., Krolik, J. H., \& Ryde, F. 1997, MNRAS, 
   292, L21
\bibitem{S97} Sincell, M. W., \& Krolik, J. H. 1997, ApJ, 476, 605
\bibitem{Z98} Zdziarski, A. A., et al.\ 1998, MNRAS, 301, 435
\bibitem{Z99} Zdziarski, A. A., Lubi\'nski, P., \& Smith, D. A., 1999,
   MNRAS, 303, L11 (ZLS)
\end{thebibliography}
\end{document}